\documentclass[a4paper,english,superscriptaddress,showpacs,pra]{revtex4}

\usepackage[utf8]{inputenc}
\usepackage{babel}
\usepackage{textcomp}
\usepackage{color}
\usepackage{amsfonts}
\usepackage{amsmath}
\usepackage{amssymb}
\usepackage{graphicx}
\usepackage{dsfont}
\usepackage{subfigure}
\usepackage{palatino}

\usepackage[normalem]{ulem} 

\DeclareMathOperator{\e}{e}

\usepackage{graphicx}

\graphicspath{{converted_graphics/}}
\begin{document}

\title{Experimental study of quantum thermodynamics using optical vortices}

\author{R. Medeiros de Ara\'{u}jo}
\affiliation{Departamento de F\'{i}sica, Universidade Federal de Santa Catarina, Florian\'{o}polis, SC, Brazil}
\author{T. H\"{a}ffner}
\affiliation{Departamento de F\'{i}sica, Universidade Federal de Santa Catarina, Florian\'{o}polis, SC, Brazil}
\author{R. Bernardi}
\affiliation{Departamento de F\'{i}sica, Universidade Federal de Santa Catarina, Florian\'{o}polis, SC, Brazil}
\author{D. S. Tasca}
\affiliation{Instituto de F\'{i}sica, Universidade Federal Fluminense, Niter\'{o}i, RJ, Brazil}
\author{M. P. J. Lavery}
\affiliation{School of Engineering, University of Glasgow, UK}
\author{M. J. Padgett}
\affiliation{School of Physics and Astronomy, University of Glasgow, Glasgow, G12 8QQ, UK}
\author{A. Kanaan}
\affiliation{Departamento de F\'{i}sica, Universidade Federal de Santa Catarina, Florian\'{o}polis, SC, Brazil}
\author{L. C. C\'{e}leri}
\email{lucas@chibebe.org}
\affiliation{Instituto de F\'{i}sica, Universidade Federal de Goi\'{a}s, Goi\^{a}nia, GO, Brazil}
\author{P. H. Souto Ribeiro}
\email{p.h.s.ribeiro@ufsc.br}
\affiliation{Departamento de F\'{i}sica, Universidade Federal de Santa Catarina, Florian\'{o}polis, SC, Brazil}

\begin{abstract}
Non-equilibrium thermodynamics and quantum information theory are interrelated research fields witnessing an increasing theoretical and experimental interest. This is mainly due to the broadness of these theories, which found applications in many different fields of science, ranging from biology to the foundations of physics. Here, by employing the orbital angular momentum of light, we propose a new platform for studying non-equilibrium properties of high dimensional quantum systems. Specifically, we use Laguerre-Gaussian beams to emulate the energy eigenstates of a two-dimension quantum harmonic oscillator having angular momentum. These light beams are subjected to a process realized by a spatial light modulator and the corresponding work distribution is experimentally reconstructed employing a two-point measurement scheme. The Jarzynski fluctuation relation is then verified. We also suggest the realization of Maxwell's demon with this platform.
\end{abstract}

\maketitle

\section{Introduction}
\noindent
The orbital angular momentum (OAM) of light is a property of the topology of the optical modes, and are characterized by discrete numbers associated to the amount of orbital angular momentum per photon in the mode \cite{Padgett}. The natural family of optical modes with orbital angular momentum are the Laguerre-Gaussian (LG) modes, a set of  solutions of the paraxial wave equation \cite{salehteich} that are described by their radial number $p$ and the azimuthal number $\ell$. The study and application of these modes is relatively recent and has increased considerably in the last two decades \cite{Padgett,molina,Babiker}.

Single photons populating modes with OAM are physical realizations of high-dimensional quantum states \cite{Vaziri,Agnew,Dada,lofller,Giovannini,Krenn1,Krenn}, leading to the possibility of encoding more than one bit of information per photon. Such photonic qudits can be explored in order to improve quantum communication schemes and quantum information processing \cite{Molina,Bourennane,Huber,Groblacher,Walborn,Mafu,Mirhosseini,Mair}. Moreover, the transverse amplitude profiles of LG light modes are formally identical to the energy eigenstates of the two-dimension quantum harmonic oscillator. Therefore, they stand as a platform for the emulation of these quantum systems in a variety of interesting problems. In the present work, we employ these light modes to experimentally study some thermodymical aspects of a high dimensional quantum system. A similar approach has been successfully used to study the quantum limits of a chaotic harmonic oscillator \cite{GBL}.  Non-equilibrium thermodynamics is fundamentally concerned to the characterization of the response of a system under external perturbations. The theory of the linear-response regime was developed in Refs. \cite{Callen,Green,Kubo}, based on earlier works such as Refs. \cite{Einstein,Johnson,Nyquist}. The information about the complete nonlinear response is contained in the so called fluctuation theorems, which have been proved for classical \cite{Bochkov,Bochkov1,Jarzynski} and for quantum systems \cite{Tasaki,Hanggi}.  

Fluctuation relations can be understood as a quantification of the probability of observing a violation of the second law of thermodynamics for small systems (when fluctuations come into play) and short time-scales. Considering the new trend in miniaturization, such fluctuations and time-scales are becoming more important for the development of new technological devices \cite{Hanggi_rev}. Therefore, the theoretical and experimental study of quantum fluctuation relations are of primary interest, both for fundamental issues and for understanding the limitations of implementing quantum information processing and communication devices. 

The quantum versions of the classical fluctuation theorems are possible only due to the two-point measurement approach for defining work. Work performed on (or by) the system is defined as the difference between two energy measurements, one before and one after the considered process takes place. To be specific, let us consider an externally driven system $\mathcal{S}$, whose time-dependent Hamiltonian is denoted by $H_{\mathcal{S}}(t)$, initially in the thermal state $\rho_{\beta}$, with $\beta=1/k_BT$, where $T$ is the temperature of the system and $k_B$ is the Boltzmann constant. The scenario considered here can be divided into three steps: $i)$ projective measurement on the initial Hamiltonian, $H_{\mathcal{S}}(0)$, eigenbasis; $ii)$ unitary (driven) evolution for a time interval $\tau$; $iii)$ projective measurement on the final Hamiltonian, $H_{\mathcal{S}}(\tau)$, eigenbasis. Defining the two-point measurement variable $W_{mn} = \varepsilon_m - \varepsilon_n$, where $\varepsilon_m$ and $\varepsilon_n$ are the eigenvalues of $H_{\mathcal{S}}(\tau)$ and $H_{\mathcal{S}}(0)$, respectively, it is not difficult to show that this stochastic variable must obey the general fluctuation relation known as the Jarzynski equality \cite{Esposito_RMP_81_1665,Campisi_RMP_83_771}
\begin{equation}
\left\langle \e^{-\beta W}\right\rangle \equiv \int d W P(W) \e^{-\beta W} = \e^{-\beta \Delta F},
\label{eq:fluc-rel}
\end{equation}
where $P(W)$ is the probability density distribution associated with the random variable $W$ and $\Delta F = F_{\tau} - F_{0}$ is the variation of the free energy over the time interval $\tau$ during which the system is subjected to the process. It has been shown that Eq. (\ref{eq:fluc-rel}) is also valid for unital processes \cite{Zyczkowsky}, i.e. quantum maps that do not change the identity. 

Note that the final state is not necessarily a thermal state, since it is generated by a projective measurement followed by an evolution. However, what appears in the right-hand side of Eq. (\ref{eq:fluc-rel}) is the equilibrium quantity $F_{\tau}$, which concerns the state the system ends up in if it is allowed to thermalize with a reservoir at the same temperature as the initial one. By defining the entropy production as $\sigma = \beta\left(W - \Delta F\right)$ we can rewrite the fluctuation relation as $\langle \e^{-\sigma} \rangle = 1$. 

The experimental investigation of such relation is new, specially in quantum systems. Regarding classical systems we can mention the experiments reported in Refs. \cite{Hummer,Liphardt,Collin,Blickle,Harris,Saira}. In the quantum regime, experiments tend to get trickier or more complicated due to the difficulty in performing energy projective measurements on arbitrary systems. Only recently, based on an alternative scheme that avoid such measurements \cite{Dorner,Mazzola}, an experimental reconstruction of the work distribution associated with a process performed on a spin-$1/2$ system was reported \cite{Batalhao}. Considering the projective measurements, the only experiment to date, as far as we know, was reported in Ref. \cite{An}, where the authors employed trapped ions in order to investigate the work statistics associated with a harmonic oscillator. Here we contribute to this line by employing the projective measurement scheme to reconstruct the probability distribution associated with a process performed on the two-dimensional harmonic oscillator with angular momentum using an optical setup, thus providing a new experimental platform for the investigation of thermodynamic processes in the quantum regime.


\section{Simulating a quantum system with classical light}

It is possible to simulate a class of quantum systems using classical light and the analogy between the paraxial wave equation and the two-dimension Schrödinger equation. This analogy has been explored experimentally to investigate the quantum limit of a chaotic harmonic oscillator \cite{GBL} and to propose a study, similar to the one done here, in which the characteristic function of the work distribution could be measured \cite{Talarico}. OAM optical modes emulate the energy eigenstates of the harmonic oscillator in the sense that the transverse distribution of the electric field of LG beams has the same form as the energy eigenfunctions of the 2-D quantum harmonic oscillator. Moreover, under appropriate conditions, the propagation of the light beams is equivalent to the Hamiltonian evolution of the harmonic oscillator \cite{GBL,Talarico,marcuse}. 

This type of simulation accounts for all oscillatory aspects of quantum systems, such as state superposition, coherence and decoherence. The intrinsic quantum properties of light itself do not come into play in this scenario, since we are exclusively interested in light's modal structure, rather than in its photonic content, which is usually explored by using detectors like avalanche photodiodes (for single photons) or low-reverse-bias photodiodes (for the continuous variables regime).

In the scheme we present here, we use the OAM modes to represent the wave functions of the two-dimension quantum harmonic oscillator, for which the Hamiltonian and angular momentum operators $H$ and $L_z$ form a Complete Set of Commuting Observables. They are written in terms of the number operators for right ($N_r$) and left ($N_l$) circular quanta as
\begin{align}
H &= (N_r + N_l + 1)\hbar\omega\\
L_z &= (N_r - N_l)\hbar
\end{align}
and their eigenvalues are
\begin{align}
&\textup{Energy: }\varepsilon_{\ell p}=(|\ell|+2p+1)\hbar\omega\\
&\textup{Angular momentum: }\lambda_\ell=\hbar\ell
\end{align}
where $\ell$ and $p$ are the azimuthal and radial \emph{quantum numbers}, respectively, which are analogous to the azimuthal and radial \emph{indices} used to identify the elements of the LG basis of modes.
Note that, for the subset of states having the quantum number $p=0$, i.e. whenever either $N_r$ or $N_l$ has eigenvalue zero, the state energy
\begin{equation}
\varepsilon_\ell = (|\ell|+1)\hbar\omega
\end{equation}
depends only on the azimuthal quantum number $\ell$. Thus, in this case, projections in the OAM basis are equivalent to projections in the energy eigenbasis.

If we restrict ourselves to processes acting on the harmonic oscillator that only change $\ell$ and we project the system's final state onto an eigenstate of the angular momentum, the work done in one experiment run depends solely on the change in $|\ell|$. Indeed, when the system goes from an initial $\ell$ to a final $\ell^\prime$, the work done is $W_{\ell \ell'} = (|\ell'| - |\ell|)\hbar\omega$. The work probability distribution is then
\begin{equation}
P(W) = \sum_{\ell,\ell'} p_{\ell\ell'}\delta\left(W - W_{\ell \ell'}\right),
\label{wdist}
\end{equation} 
where $p_{\ell\ell'} = p_{\ell}\ p_{\ell'|\ell}$ is the probability of observing the transition $\ell\rightarrow\ell'$, with $p_{\ell}$ being the probability of having $\ell$ at the input and $p_{\ell'|\ell}$ the probability of observing $\ell'$ at the output given that the input is $\ell$.

In the context of Jarzynski equality, $p_{\ell}$ is found in the expression of the initial thermal state $\rho_{\beta} = \e^{-\beta H}/Z$, where $Z$ is the partition function. This state may be explicitly written as:
\begin{equation}
\rho_\beta=\sum_{\ell=-\infty}^{+\infty} p_\ell |\ell\rangle\langle\ell|,\quad\textup{with}\quad p_\ell=\frac{\mbox{e}^{-\beta\varepsilon_\ell}}{Z}\quad \textup{and}\quad Z=\e^{\beta\hbar\omega}\tanh\frac{\beta\hbar\omega}{2}.
\label{eq:thermal}
\end{equation}
Note that, since the states $|\ell\rangle$ and $|-\ell\rangle$ have the same energy, their probabilities are the same: $p_\ell = p_{-\ell}$. In fact, every energy level has degeneracy 2, except for the ground state $|\ell=0\rangle$.

\section{Experimental setup}
\label{exp}

\begin{figure*}[t]
\begin{subfigure}{}
\includegraphics[width=0.6\textwidth]{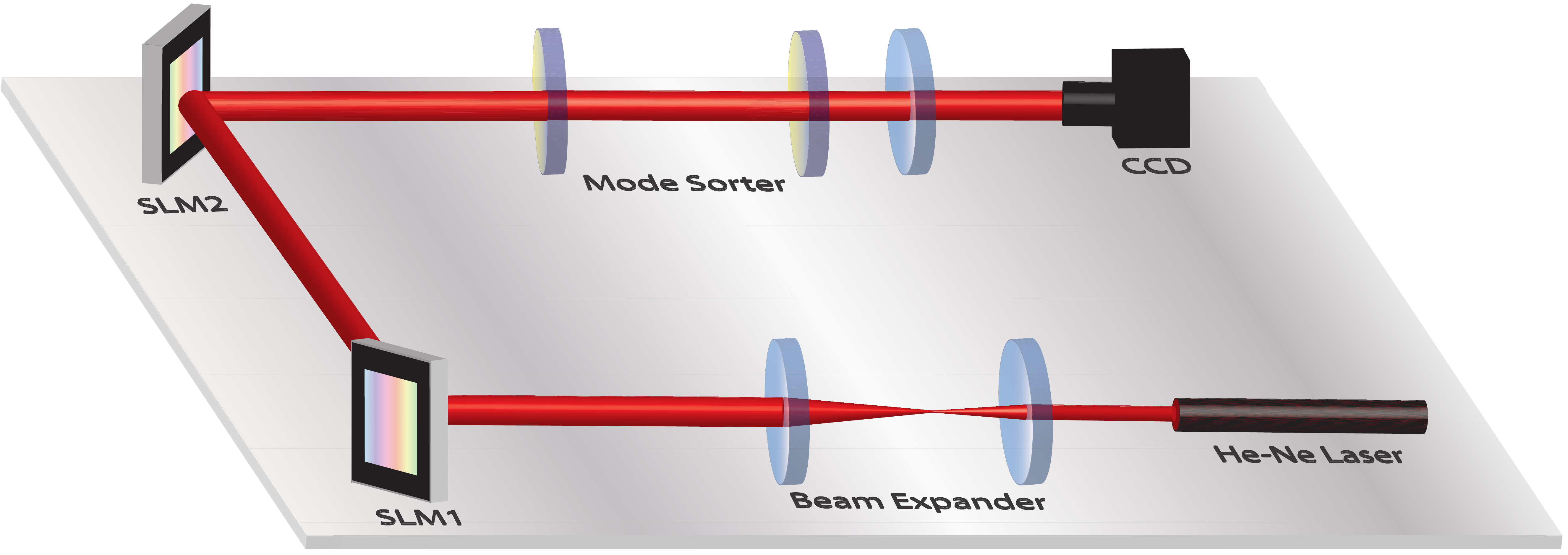}
\end{subfigure}
\begin{subfigure}{}
\includegraphics[width=0.3\textwidth]{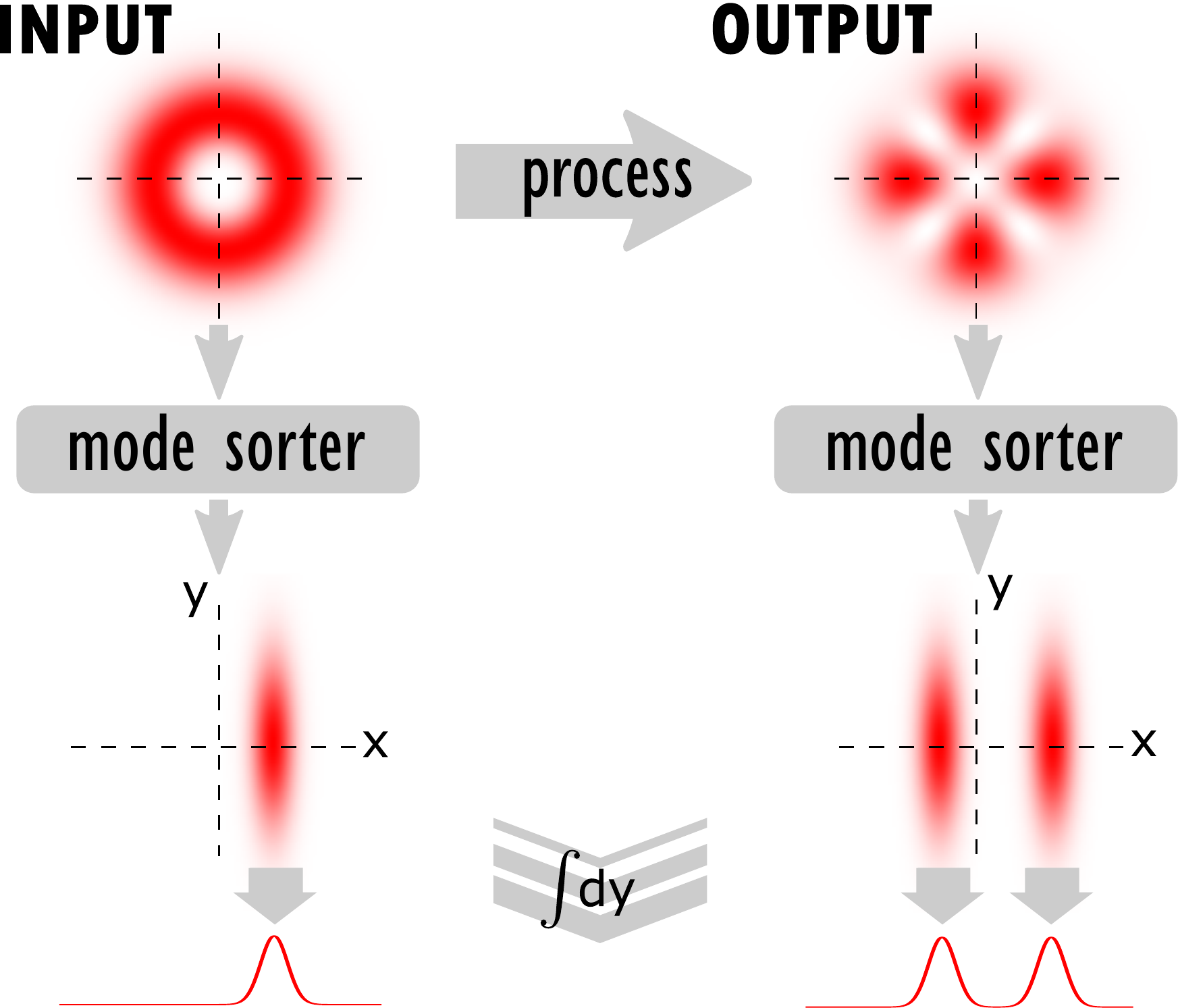}
\end{subfigure}
\caption{Experimental setup (left): SLM1 generates an input OAM mode, which undergoes a process implemented by SLM2. Its output is analysed by a mode sorter. General idea (right): the mode sorter sorts the OAM components along the $x$-axis of the CCD camera. The image is then integrated along the $y$-axis.}
\label{fig:setup}
\end{figure*}

The sketch of the experimental setup is shown in Fig. \ref{fig:setup}. The light from a He-Ne laser is sent through a beam expander consisting of two lenses in a confocal arrangement, with focal lengths $f_1 = 50$ mm and $f_2 = 300$ mm, resulting in an expansion factor of 6. The expanded beam is sent to the first Spatial Light Modulator (SLM1), where an OAM mode is prepared with the usual approach with a forked hologram \cite{Padgett}. 


\begin{figure}
\centering
\includegraphics[width=0.5\columnwidth]{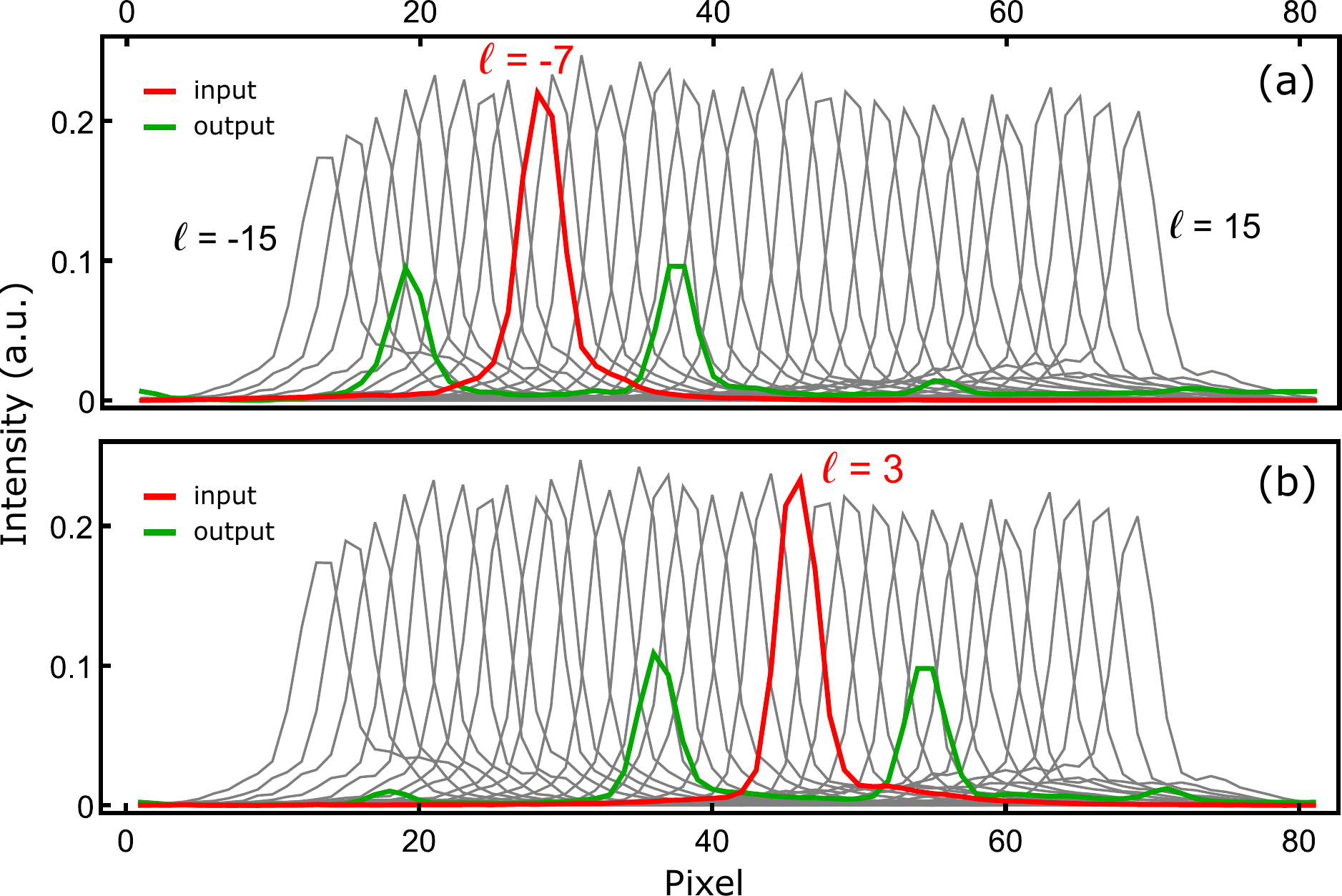}
\caption{Intensity distributions at the output of the mode sorter.  In gray, the calibration curves obtained by sending OAM modes ranging from $\ell=-15$ to $\ell=+15$, with no process applied (flat phase mask on SLM2). Colored curves: (a) process $(\mathcal{L}_{+5} + \mathcal{L}_{-5})/\sqrt{2}$ (defined from Eq. \ref{eq:ellpm}) is applied by SLM2 onto $\ell = -7$, splitting the input mode into two modes with $\ell'=-12$ and $\ell'=-2$; (b) Input at $\ell = 3$ split up into $\ell'=-2$ and $\ell'=8$ by the same process. Each curve is obtained by integrating the output intensity profile over the vertical direction of camera.}
\label{fig:curves}
\end{figure}

SLM1 generates modes with OAM $\hbar\ell$ per photon and they are sent to a second spatial light modulator, SLM2, where another phase mask realizes some operation on them depending on the protocol. The resulting light beam is sent to a device called \emph{mode sorter} \cite{modesorter}, which litterally sorts the different OAM components of the beam along the horizontal axis of a CCD camera. The measurement scheme is calibrated sending OAM modes one by one and measuring the intensity distribution at the output of the mode sorter with no phase modulation on SLM2 (flat phase mask). Typical calibration curves are shown in gray in Fig. \ref{fig:curves}, where each curve represents the intensity distribution at the output for a given value of $\ell$, from -15 to +15 in this case.
 
Step \emph{i} of the two-point measurement protocol consists in preparing the thermal state described by Eq. (\ref{eq:thermal}) and performing a projective measurement in the initial Hamiltonian eigenbasis or, equivalently, in the OAM basis. 

The preparation of the thermal state is made by sending a Gaussian mode with $\ell = 0$ to the spatial light modulator SLM1, which applies masks that generate OAM states with $\ell$ ranging from $-7 \leq \ell \leq 7$. Each mask is turned on for 3 s, according to a random choice of $\ell$ with weight $p_\ell$. The resulting light beam is sent to SLM2, that acts just as mirror in this case, and then to the  mode sorter, which analyses the OAM components. The light intensity at the output of the mode sorter is measured with a CCD camera, and the images are analyzed as explained in detail in Appendix A. A typical result is shown in Fig. \ref{fig:thermal}, where the final distribution is obtained from 300 runs of the experiment. The distribution obtained for the \emph{absolute value} of OAM is normalized and fitted to the function 
\begin{equation}
p(|\ell|)=N\textup{e}^{-\beta\varepsilon_{|\ell|}}=N\textup{e}^{-\beta(|\ell|+1)\hbar\omega},
\end{equation}
which represents the Boltzmann Distribution, leaving $N$ (the normalization factor) and $\beta\hbar\omega$ as free parameters. For the results shown in Fig. \ref{fig:thermal}, we obtained an excellent agreement with the fitted function, with $\beta\hbar\omega$ = 0.67 $\pm$ 0.01.

The parameter $\beta\hbar\omega=\hbar\omega/k_BT$ can be interpreted as the ratio between the ground state energy and the typical scale of thermal energy at temperature $T$. So, for a given system, the greater $\beta\hbar\omega$, the lower the temperature.

We show how to prepare a thermal state and perform the projective measurements in the energy eigenbasis in order to illustrate this procedure. However,  in the second step of the protocol, we prepare each energy eigenstate separately and submit it to the process in order to obtain the transition probabilities. This is strictly equivalent to using the states resulting from the first measurement and means a considerable simplification in the set up.

\begin{figure}
\includegraphics[width=0.5\textwidth]{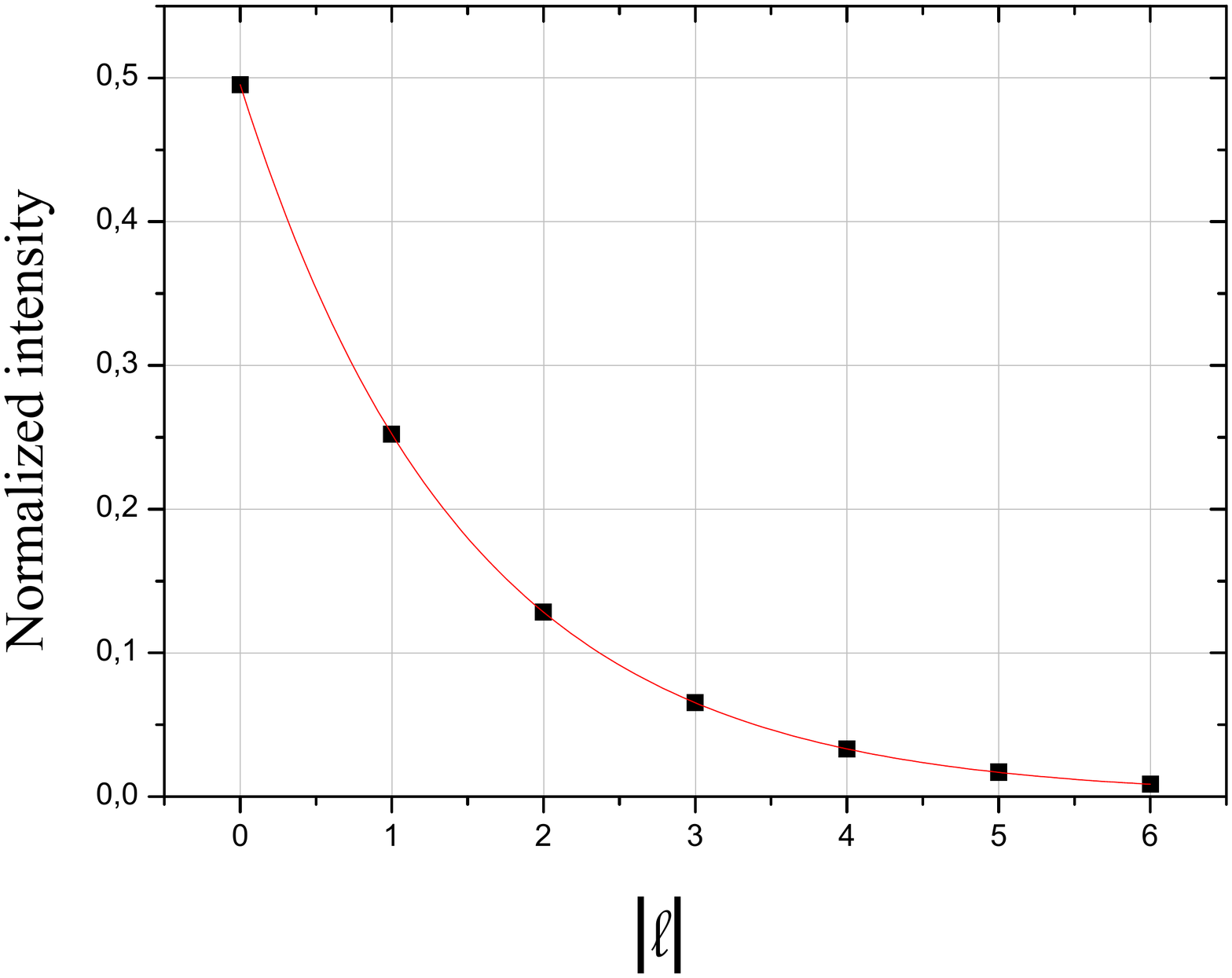}
\caption{Normalized intensity distribution as a function of $|\ell|$. $\beta\hbar\omega$ = 0.67 $\pm$ 0.01.}
\label{fig:thermal}
\end{figure}

Step \emph{ii} consists in sending input modes prepared with SLM1 and having OAM ranging from $\ell = -7$  to $\ell = +7$ to SLM2, where a phase mask is applied, realizing the process whose work distribution will be measured. This mask couples the input mode to other OAM modes, thus inducing OAM transitions. The energy spectrum of the system is discrete and infinite. However, the thermal weight of states corresponding to higher energies can be made negligible by choosing sufficiently low temperatures, so that we can truncate the initial distribution of states, as we did.

Using the mode sorter, we are able to measure the final distribution of OAM modes and their corresponding weights. This device implements a projective measurement in the orbital angular momentum basis, which in our case is equivalent to the energy under the assumption that the radial number $p = 0$. Observing the mode sorter output, we can compute the transition probabilities and, consequently, reconstruct the work distribution. Nonetheless, as the calibration figure shows, there is a considerable overlap between adjacent curves (adjacent orbital angular momenta). This appreciably reduces the resolution of the OAM sorter. Newer generations of mode sorters, as well as other strategies \cite{resolve}, minimise this technical inconvenient.

In the present proof of principle experiment, we overcome this issue using a process in SLM2 which generates superpositions of OAM modes that can be easily resolved by the mode sorter. Specifically, the process in our experiment implements the linear operation $(\mathcal{L}_{+5} + \mathcal{L}_{-5})/\sqrt{2}$, where we define
\begin{equation}
\mathcal{L}_{\pm5}|\ell\rangle = |\ell\pm 5\rangle.
\label{eq:ellpm}
\end{equation}
In this way, the overlap between the two components at the output becomes negligible. Typical measurement results are shown in Figs. \ref{fig:curves}(a) and \ref{fig:curves}(b).

\begin{figure*}[tbp]
\includegraphics[width=0.8\textwidth]{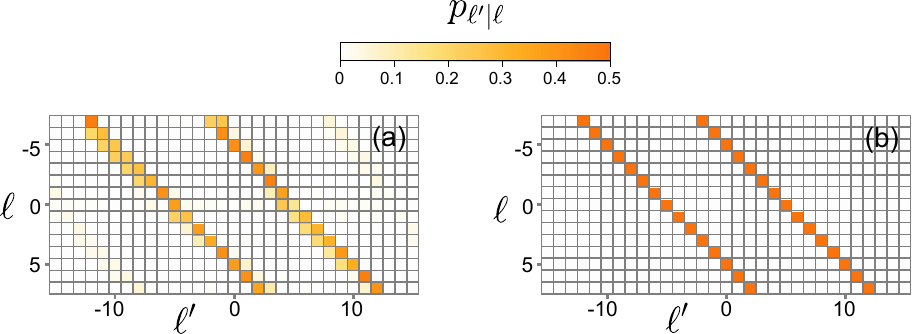}
\caption{Conditional transition probabilities $p_{\ell |\ell'}$. (a) Input-output matrix obtained from the experimental results for the process $(\mathcal{L}_{+5} + \mathcal{L}_{-5})/\sqrt{2}$ applied to input modes $-7 \leq \ell \leq 7$. (b) Theoretical prediction for the same process. }
\label{fig:matrix}
\end{figure*}

For each measured output, we performed a linear least squares regression in order to obtain the values of the orbital angular momenta and their respective weights (see Appendix A for details). The normalised set of all OAM weights for all outputs is exhibited in the matrix of Fig. \ref{fig:matrix}(a). This is a density plot where the index of the 
input (output) modes are the labels of the vertical (horizontal) axis. In other words, these are the transition probabilities for a typical run of the experiment. Fig. \ref{fig:matrix}(b) shows the matrix for the ideal process.

\section{Results and Discussion}

In our experiment, we measured the conditional transition probabilities $p_{\ell'|\ell}$ as shown in Fig. \ref{fig:matrix}. As explained earlier, we truncated the OAM space and limited the input modes to $|\ell|\leq 7$. That is to say we operate in a regime of low temperatures where the Boltzmann weights for $|\ell|>7$ can be neglected, i.e. $\beta\hbar\omega\gtrsim 1$, i.e. $\hbar\omega\gtrsim k_BT$.

\begin{figure}[h]
\includegraphics[width=\columnwidth]{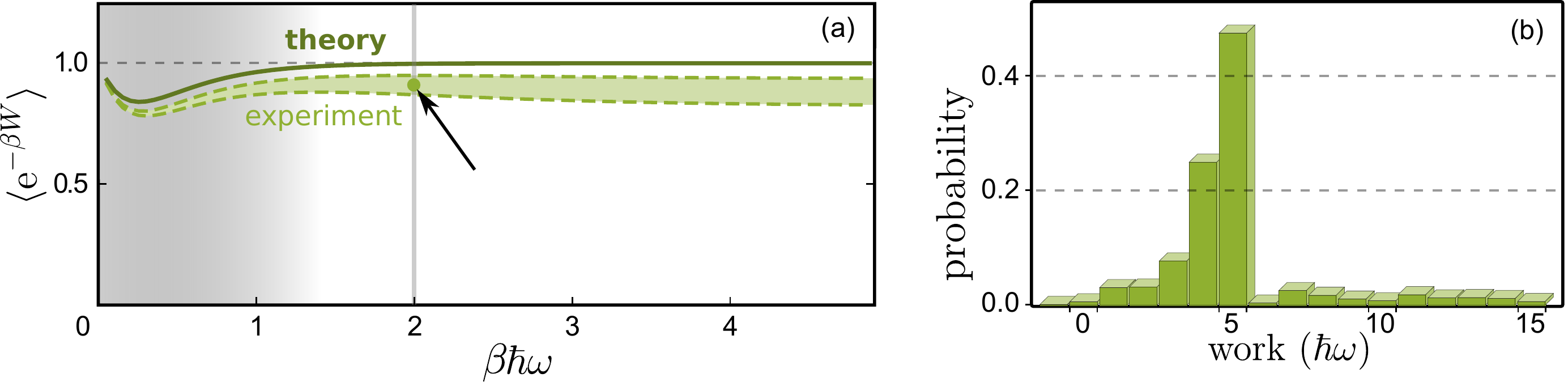} 
\caption{\textbf{(a) Fluctuation relation}. Plot of  $\left\langle\e^{-\beta W} \right\rangle$ for the process  $(\mathcal{L}_{+5} + \mathcal{L}_{-5})/\sqrt{2}$. The shaded gray area indicates the region where the effect of truncation of the input states is non negligible. Curves labelled {\em theory} and {\em experiment} are obtained using an ideal process and the experimental results, respectively. The filled region between the dashed lines (experimental curves) represents the assessed measurement uncertainty, within a 95\% confidence level (see Appendix B for details). \textbf{(b) Work distribution}. Experimentally reconstructed probability distribution for each possible value of work with $\beta\hbar\omega = 2$ (point indicated by the arrow at subfigure (a)).}
\label{fig:jar}
\end{figure}

Fig. \ref{fig:jar} shows a plot of the quantity $\left\langle\e^{-\beta W}\right\rangle$ as a function of $\beta\hbar\omega$. The inset displays the probability distributions of work computed from the measurement results for $\beta\hbar\omega = 2$. The probabilities in the vertical axis are obtained summing up all values of $p_{\ell\ell'}$ (given in Eq.\ref{wdist}) for which the corresponding transition results in a given value of work $W$.  Regarding the Jarzynski equality, Eq. (\ref{eq:fluc-rel}), the considered process gives $\Delta F = 0$ as it does not change the Hamiltonian of the system. In other words, transitions are induced, but the energy levels do not change. Therefore, Eq. (\ref{eq:fluc-rel}) becomes simply $\left\langle\e^{-\beta W}\right\rangle = 1$. 

The curve named {\em theory} represents $\left\langle\e^{-\beta W}\right\rangle$ computed for an ideal (but truncated) process. The corresponding ideal transition probabilities are illustrated in Fig. \ref{fig:matrix}(b). The curve named {\em experiment} represents the same quantity computed from the measured transition probabilities shown in Fig. \ref{fig:matrix}(a). Within the range $\beta\hbar\omega\lesssim 1$ (gray area), we obtain $\left\langle\e^{-\beta W} \right\rangle < 1$  even for the {\em theory} curve, due to the truncation of the input states in $|\ell| \leq 7$. However, for larger values of $\beta\hbar\omega$, we see that the {\em theory} curve is essentially constant and equal to 1, while the {\em experiment} curve is always below unity. We interpret this result as a consequence of entropy increase due to natural technical limitations present in a real world measurement (which includes classical fluctuations coming from laser pointing instability, mechanical vibrations on the set-up and camera thermal noise) and experimental imperfections (such as misalignment, limited pixel resolution on the SLM and on the camera and limited optical resolution on the mode sorter). The uncertainty band shown in Fig. \ref{fig:jar} accounts only for the fluctuations detected upon several subsequent identical measurements (see Appendix B for details). For instance, for $\beta\hbar\omega=2$, we have found $\e^{-\beta W}=0.910\pm 0.046$ (95\%-confidence interval), clearly different from 1. We atribute the remaining difference to the experimental imperfections listed above that are not captured by our error estimation procedure, but that is captured by the Jarzynski's fluctuation relation.

\section{Maxwell's demon}

While Eq. (\ref{eq:fluc-rel}) is valid for any unital process \cite{Zyczkowsky}, in a general context including measurements and feedback, i.e., when Maxwell's demon come into play, a new equality holds \cite{Ueda}:
\begin{equation}
\left\langle \e^{-\sigma - I}\right\rangle = 1,
\label{ueda}
\end{equation}
where $I$ is the information acquired by the demon due to the measurement process. This implies a modification of the second law of thermodynamics as $\sigma \geq -I$, highlighting the demon's main feature, which is the use of information to reduce entropy production.

\begin{figure}[h]
\includegraphics[width=\columnwidth]{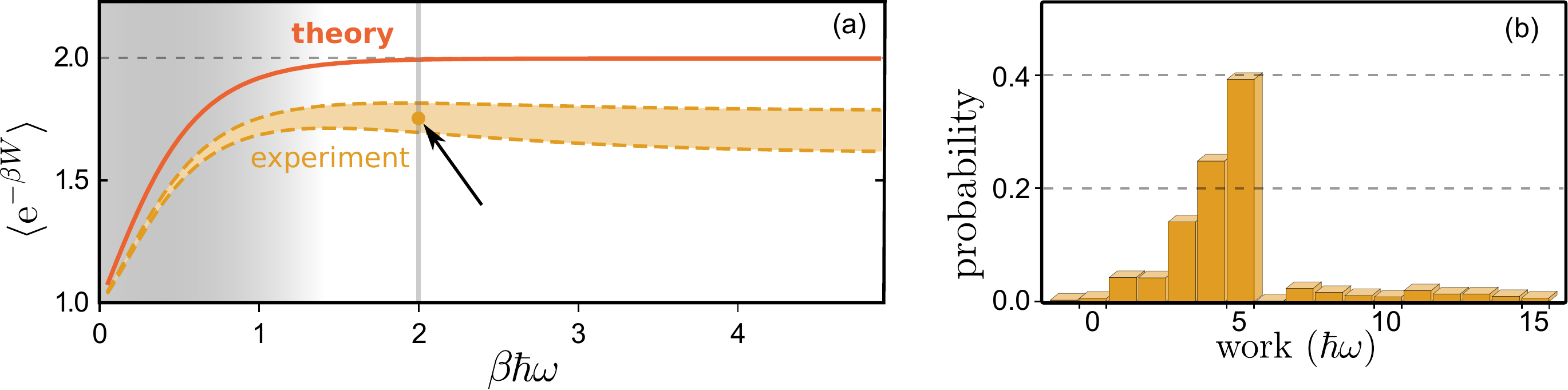} 
\caption{\textbf{(a) Fluctuation relation}. Plot of  $\left\langle\e^{-\beta W} \right\rangle$ for the Maxwell's demon scheme with process  $(\mathcal{L}_{+5}$ applied to negative OAM modes and $\mathcal{L}_{-5}$ applied to negative OAM modes. The shaded gray area indicates the region where the effect of truncation of the input states is non negligible. Curves labelled {\em theory} and {\em experiment} are obtained using an ideal process and the results from a \emph{simulated} experiment, respectively. \textbf{(b) Work distribution}. Experimentally reconstructed probability distribution for each possible value of work with $\beta\hbar\omega = 2$ (point indicated by the arrow at subfigure (a)).}

\label{fig:demon}
\end{figure}

\begin{figure}[h]
\includegraphics[width=0.5\columnwidth]{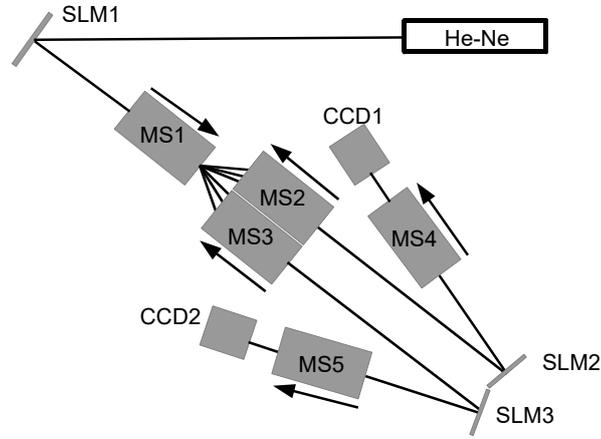} 
\caption{\textbf{Sketch of the scheme}. He-Ne is a Helium-Neon laser. SLM is spatial light modulator. MS is mode sorter. CCD is charge coupled device, which is a camera. The arrows near the mode sorters indicates in which sense they are being used.}
\label{MD}
\end{figure}

We propose an experimental scheme for realizing Maxwell's demon using OAM modes. Fig. \ref{MD} shows the sketch of the suggested scheme. A laser beam is sent to spatial light modulator SLM1, which produces a thermal state of OAM modes as discussed in Sec. \ref{exp}. The light beam prepared in the thermal state is then sent through mode sorter MS1. The modes with positive angular momentum will be deflected to the right, while those with negative angular momentum will drift to the left. These two groups of beams are separated and directed to mode sorters MS2 and MS3 working in reverse \cite{reverse1,reverse2}, converting them back to OAM modes. With this approach, it is possible to separate OAM modes according to sign of $\ell$. SLM2 is used to apply the operation  $\mathcal{L}_{+5}$ to the modes with negative $\ell$ and SLM3 is used to apply $\mathcal{L}_{-5}$ to the modes with positive $\ell$. They are finally sent to mode sorters MS4 and MS5 that perform the final measurement of the OAM using CCD cameras CCD1 and CCD2. The measurement and feedback, in this case, increases the probability of lowering the absolute value of orbital angular momentum ($|\ell|$) of the system, thus extracting work from the initial thermal state. As an example, let us start with $\ell=+3$: the operation $(\mathcal{L}_{+5} + \mathcal{L}_{-5})/\sqrt{2}$, and subsequent measurement, results in either $\ell=-2$ or  $\ell=+8$ with equal probabilities, leading to an average work of $2\hbar\omega$. Now, if Maxwell's demon takes action, $\ell$ goes invariably from +3 to -2 and $W=-\hbar\omega<0$, which means that work is extracted from the system.

As the measurement of OAM sign provides Maxwell's demon with one bit of information, $I=\ln 2$, and Eq. (\ref{ueda}) leads to $\left\langle \e^{-\sigma}\right\rangle = 2$ for all $\beta\hbar\omega$, which is the Jarzynski's fluctuation relation with demon's action for the ideal case. This scenario has been computed and plotted in Fig. \ref{fig:demon} (curve named \emph{theory}). Note that the values of $\left\langle\e^{-\sigma} \right\rangle$ below 2 at high temperatures (shaded gray area on the left) are simply due to truncation of the OAM space dimension.

We would also like to have some insight on the effect of noise in the Maxwell's demon scenario. In order to do that, we have added the same noise that appears in Fig. \ref{fig:matrix} to the probability transitions of the Maxwell's demon scheme. The result is the curve named \emph{noise} in Fig. \ref{fig:demon}, that carries an uncertainty band calculated in the same way as for the experimental curve in Fig. \ref{fig:jar}. Notice that the random noise decreases the average value of $\e^{-\sigma}$, meaning that the decrease in entropy production caused by Maxwell's demon would be affected by experimental noise.

The inset shows the probability distributions for the possible values of work in the case of $\beta\hbar\omega =$ 2. Even though the value of $\left\langle\e^{-\sigma} \right\rangle$ changes dramatically from 1 to 2 when the demon takes action, the work distribution and the average work do not change too much: $\langle W\rangle = 5.0\,\hbar\omega$ without Maxwell's demon and $\langle W\rangle = 4.8\,\hbar\omega$ with it. This is due to the fact that, in this sytem, for low enough temperatures such as $\beta\hbar\omega = 2$, there is a large contribution from the state component with $\ell =0$  in the input thermal state. For an input with $\ell = 0$, both $\mathcal{L}_{+5}$ and $\mathcal{L}_{-5}$ contribute for positive work, which dominates the work distribution.

\section{Conclusion}

In conclusion, we have experimentally investigated the quantum version of thermodynamic work and the Jarzynski's fluctuation relation using the orbital angular momentum of light, a discrete degree of freedom with infinite dimension usually employed in the single-photon regime to realize a qudit, with applications in quantum communication and quantum information processing. Here, by exploring the analogy between the Paraxial wave equation and Schrödinger equation, we have used OAM of light to simulate the eigenstates of the two-dimensional quantum harmonic oscillator and their evolution through a given process. We measured the work distribution associated with this process and obtained the experimental Jarzynski's fluctuation relation. We have also proposed a scheme for implementing a Maxwell's demon measurement and feedback action. These results illustrate the usefulness of Laguerre-Gaussian beams as a practical platform to investigate aspects of the growing field of quantum thermodynamics in high dimensional Hilbert spaces. Given the versatility of this platform, one can consider using it, for example, in the study of the role of multipartite entanglement in thermodynamic processes, as well as the role of the environment, i.e., non-unitary and non-unital processes.

\section{Acknowledgments}

We acknowledge financial support from the Brazilian funding agencies CNPq, CAPES, and the National Institute for Quantum Information (INCT-IQ).
M.P.J.L. acknowledges support from EPSRC awards  EP/N032853/1.
M.J.P. acknowledges support from EPSRC QuantIC (EP/M01326X/1) and ERC TWISTS (Grant No. 192382).

\section{Appendix}

\subsection{Data fitting}
\label{fit}

When an Laguerre-Gaussian mode passes through a mode sorter, its intensity profile (initially presented in a donut-like shape) becomes an elongated spot along the vertical direction on the camera. The integration of such an image along the vertical axis gives a curve (among the 31 calibration curves on Figs. 2(a) and 2(b)) showing the marginal distribution of intensity along the horizontal direction of the sorted LG mode. The right side of Fig. 1 outlines this idea.

The horizontal size of an image on the camera ($\sim$ 80 pixels) sets the extension of the horizontal axis in Fig. 2, which shows collections of lists of 80 elements. Let us call the lists corresponding to the LG inputs $\mathbf{x}_j = \{x_{j,k}\}_{0\leq k<80}$ and the lists at the output $\mathbf{y}_i = \{y_{i,k}\}_k$, where $-7\leq i \leq 7$ and $-15\leq j\leq 15$ may be associated with $\ell$ and $\ell^\prime$, respectively. We model the process implemented by SLM2 as a linear operation represented by a $15\times 31$ matrix $\mathbf{A}$ acting on the set of possible input modes $\mathbf{X} = \{\mathbf{x}_i\}_i$ and leading to the set of its outcomes $\mathbf{Y} = \{\mathbf{y}_j\}_j$, i.e.
\begin{equation}
\mathbf{Y} = \mathbf{A} \mathbf{X}.
\label{eq:linear}
\end{equation}

In order to find the matrix that best fits our experimental data, we apply a linear least squares approach, numerically solving the minimization problem
\begin{equation}
\min_\mathbf{A}||\mathbf{Y} - \mathbf{A}\mathbf{X}||^2
\end{equation}
{\em with the additional constraints that all elements of matrix $\mathbf{A}$ must be non-negative and that the sum of the elements in each line must equal 1}. The result is a matrix similar to the ones shown in Fig. 3.

The non-negativity is equivalent to the assumption that the overall phases of each OAM component of the output are the same or, at least, that the OAM components are far enough from each other so that they do not interfere and their relative phases do not play a significant role. The sum over each line equaling 1 stands for the unitarity of the process (no optical loss), which can be assumed without loss of generality. 

\subsection{Measurement uncertainty}
\label{uncertainty}

A {\em measurement} consists in taking a picture, integrating it and obtaining its marginal intensity distribution $\mathbf{y}_i=\{y_{i,k}\}_k$. The goal here is to assess the uncertainty on each $y_{i,k}$ measured and, from this, to calculate the uncertainty:
\begin{itemize}
\item on each element of the matrix of conditional transition probabilities of Fig. 3(a) and
\item on the mean value $\langle\e^{-\beta W}\rangle$ as a function of $\beta\hbar\omega$ (uncertainty band on Fig. 4).
\end{itemize}

Let us call $\sigma_{i,k}$ the uncertainty on $y_{i,k}$. In order to assess these $\sigma$'s, we performed a series of ten identical measurements on the same transformed mode over a time window of a few minutes and obtained a set of intensity distributions fluctuating, for each $k$, around a mean value $\mu_k$ with standard a deviation $\sigma_k$. These standard deviations turn out to be dependent on $\mu_k$ and on $k$ itself, but mostly on $\mu_k$. We noticed, for instance, that the relative standard deviation ($\sigma/\mu$) is always smaller than 10\%, for any $k$. These measurements were used to model the typical error associated to a $\mathbf{y}_i$ measurement. This procedure led to a set of numbers $\sigma_{i,k}$ used as input for our model, in which we assume each $\mathbf{y}_i$ measured is a realization of a random variable following the multivariate normal distribution with estimated mean values $y_{i,k}$ and standard deviations $\sigma_{i,k}$. This procedure above allows us to simulate sets of measurements, realizing {\em Monte Carlo experiments}.

Ten different experimental matrices $\mathbf{Y}$ were randomly generated in this manner, from each of which we numerically solved the minimization problem above to find a different probability matrix $\mathbf{A}$. We could see from the set os matrices $\mathbf{A}$ that the relative uncertainty on each matrix element was never bigger than 2\%.

Similarly (and finally), we performed 1000 Monte Carlo experiments in order to estimate the uncertainty on $\langle\e^{-\beta W}\rangle$ for each $\beta\hbar\omega$ ranging from 0.05 to 5. We observed that the random variable $\e^{-\beta W}$ nearly follows a normal distribution for all values of $\beta\hbar\omega$. For instance, for $\beta\hbar\omega=2$, we have found $\langle\e^{-\beta W}\rangle=0.910$ with a standard deviation $\sigma=0.022$. From the $1.96\sigma$ rule, we established our 95\%-confidence interval for $\e^{-2W/\hbar\omega}$ to be $0.910\pm0.046$. By doing that same for all values of $\beta\hbar\omega$, we were able to plot the uncertainty band shown in Fig. 4.

\end{document}